\documentclass[twocolumn]{article}

\usepackage{amsmath}                    
\usepackage{graphicx}                   
\usepackage{verbatim}                   
\usepackage{color}                      
\usepackage[utf8]{inputenc}             

\usepackage[separate-uncertainty = true, multi-part-units = single]{siunitx}    
\usepackage[ngerman, english]{babel}              

\DeclareUnicodeCharacter{2212}{-} 

\usepackage[symbol*]{footmisc}
\DefineFNsymbolsTM{otherfnsymbols}{%
  \textasteriskcentered *
  \textsection   \mathsection
  \textdagger    \dagger
  \textdaggerdbl \ddagger
  \textbardbl    \|%
  \textparagraph \mathparagraph
}%

\setfnsymbol{otherfnsymbols}

\usepackage{authblk}                    
\newenvironment{acknowledgements}
{\section*{Acknowledgements}}
{}

\author[1]{Giovanni Piredda \thanks{giovanni.piredda@fhv.at}}
\author[1]{Sandra Stroj}
\author[2]{Dorian Ziss}
\author[2]{Julian Stangl}
\author[3]{Rinaldo Trotta}
\author[4]{Javier Mart\'{i}n-S\'{a}nchez \thanks{javiermartin@uniovi.es}}
\author[2]{Armando Rastelli}
\affil[1]{Josef Ressel Center for material processing with ultrashort pulsed lasers, Research Center for Microtechnology, Vorarlberg University of Applied Sciences, Hochschulstra{\ss}e 1, Dornbirn, 6850, Austria}
\affil[2]{Institute of Semiconductor and Solid State Physics, Johannes Kepler University, Altenbergerstra{\ss}e 69, Linz, 4040, Austria}
\affil[3]{Dipartimento di Fisica, Sapienza Universit\'{a} di Roma, Piazzale A. Moro 1, I-00185 Roma, Italy}
\affil[4]{Departamento de Física, Universidad de Oviedo, Oviedo, Spain}
\date{}                     
\begin{document}

\title{Micromachining of PMN-PT Crystals with Ultrashort Laser Pulses}

\date{\today}

\begin{abstract}
Lead-magnesium niobate lead-titanate (PMN-PT) has been proven as an excellent material for sensing and actuating applications. The fabrication of advanced ultra-small PMN-PT-based devices relies on the availability of sophisticated procedures for the micro-machining of PMN-PT thin films or bulk substrates. Approaches reported up to date include chemical etching, excimer laser ablation and ion milling. To ensure an excellent device performance, a key mandatory feature for a micro-machining process is to preserve as far as possible the crystalline quality of the substrates; in other words, the fabrication method must induce a low density of cracks and other kind of defects. In this work, we demonstrate a relatively fast procedure for the fabrication of high-quality PMN-PT micro-machined actuators employing green femtosecond laser pulses. The fabricated devices feature absence of extended cracks and well defined edges with relatively low roughness, which is advantageous for the further integration of nanomaterials onto the piezoelectric actuators.
\end{abstract}

\maketitle

\section{Introduction}

Piezoelectric materials offer interesting properties for micro-electro-mechanical systems (MEMS) including sensing and actuating applications, as they provide wide dynamic ranges, low power consumption and low hysteresis values \cite{Holterman_2013_PiezoelectricMaterialsandApplications}. Among them, lead-magnesium niobate lead-titanate (PMN-PT) crystals represents an ideal material for the fabrication of actuators due to its giant piezoelectric response \cite{Park_1997_UltrahighStrainandPiezoelectricBehaviorinRelaxorBasedFerroelectricSingleCrystals,Baek_2012_GiantPiezoelectricityinPMNPTThinFilmsbeyondPZT}. In particular, piezoelectric actuators fabricated by femtosecond laser cutting have been recently demonstrated for tailoring the optical properties of nanomaterials \cite{Martin-Sanchez_2018_StrainTuningoftheOpticalPropertiesofSemiconductorNanomaterialsbyIntegrationontoPiezoelectricActuators,Martin-Sanchez_2016_ReversibleControlofinPlaneElasticStressTensorinNanomembranes}; these devices are based on the inverse piezoelectric effect where the in-plane deformation state (strain) of a piezoelectric substrate upon application of an out-of-plane electric field is transferred to a medium bonded onto it. The practical implementation of such devices depends on the development of adequate procedures to machine PMN-PT thin films or substrates with strict requirements like spatial resolution within the micrometric range, sharp edges, relatively low roughness to allow fine control of small gaps between features, cleanliness to allow for subsequent bonding and metallization steps and quick fabrication to allow for repeated device development/testing cycles. The patterning of PMN-PT substrates has been performed by different means like wet etching \cite{Peng_2008_MicroPatterningof070PbMg13Nb23O3030PbTiO3SingleCrystalsbyUltrasonicWetChemicalEtching}, dry etching \cite{Ivan_2012_PMNPTleadMagnesiumNiobateLeadTitanatePiezoelectricMaterialMicromachiningbyExcimerLaserAblationandDryEtchingDRIE}, reactive ion etching \cite{Zhang_2015_DeepReactiveIonEtchingofPZTCeramicsandPMNPTSingleCrystalsforHighFrequencyUltrasoundTransducers}, ion milling \cite{Baek_2012_GiantPiezoelectricityinPMNPTThinFilmsbeyondPZT,Chen_2017_AddressableandColorTunablePiezophotonicLightEmittingStripes} and nanosecond UV laser ablation \cite{Ivan_2012_PMNPTleadMagnesiumNiobateLeadTitanatePiezoelectricMaterialMicromachiningbyExcimerLaserAblationandDryEtchingDRIE,Lam_2013_KerfProfileandPiezoresponseStudyoftheLaserMicroMachinedPMNPTSingleCrystalUsing355nmNdYAG}.

The latter introduces roughening of the edges on the scale of about \SI{10}{\micro\meter} and consequently low definition in their shapes. The rest of the approaches are based on the use of lithography techniques involving the use of masks or relatively complicated processes that make the whole process slow; this is especially remarkable when relatively thick substrates (\SIrange{200}{300}{\micro\meter}) have to be cut through; for these kind of substrates wet etching processes in particular achieve limited steepness of the side walls, which in turn limits the ability to fabricate devices with small gaps of high aspect ratio. Moreover, since PMN-PT substrates are brittle, simple and maskless fabrication procedures are desirable.
Laser microfabrication with ultrashort pulses has been established in the last ten years as a
reliable technique which allows precise material ablation with very small effects on the material
surrounding the ablation zone \cite{Sugioka_2010_LaserPrecisionMicrofabrication}. Several characteristics of ultrashort-pulse microfabrication are relevant to the preparation of PMN-PT devices. The smoothness of cut edges will prevent crack propagation and the low angle and smoothness of cut walls will allow the preparation of devices with small gaps between features. In addition, the fabrication does not involve chemical substances, whose residuals could be too aggressive for materials used in further processing steps, and the process is flexible so that devices with complex shapes can be easily fabricated.
In this work, we study the micro-machining of 300-micrometer-thick PMN-PT substrates by employing amplified laser systems providing pulses with an energy of several \si{\micro\joule} and pulse duration of about \SI{380}{\femto\second} at the wavelength of \SI{520}{nm}. The processing parameters are optimized to obtain micro-machined PMN-PT substrates with well-defined edges and arbitrary shapes. We measure ablation rates for a range of parameters and delimit the parameter region in which no heat effects occur; we examine the formation of cracks at the cut edges and the smoothness and angle of the cut walls.

\section{Materials and experimental setup}

500 micron-thick $\mathrm{PMN_{0.71}-PT_{0.29}}$ crystals were purchased from TRS Technologies, State College, Pennsylvania, U.S.A (from now on referred to as TRS crystals). They were lapped down to the thickness of \SI{300}{\micro\meter} and mechanically polished. For the ablation and cutting tests we used a femtosecond laser (Spirit\textsuperscript{\tiny\textregistered}, Spectra-Physics)  with a pulse energy of\SI{380}{\femto\second} and a pulse repetition rate of \SI{200}{\kilo\hertz}, adjustable with the use of an acusto-optical pulse picker. The laser was operated at its second harmonic with a wavelength of \SI{520}{\nano\meter} and was focused by a \SI{100}{\milli\meter} f\nobreakdash-theta telecentric objective onto the sample surface to a spot with a \SI{7}{\micro\meter} radius. For x-ray diffraction (XRD) analysis, a PANalytical X'Pert Pro MRD diffractometer was used. The high resolution omega-2theta line-scans where performed with with an hybrid monochromator in the primary beam path and an Ge[220] analyser-crystal in combination with a Xenon proportional counter in the secondary beam-path. For reciprocal space mapping the diffractometer was equipped with a 1D PIXcel detector in combination with a Soller-slit in the secondary beam-path. For all measurements we used copper k-alpha radiation, \SI{8048}{\electronvolt}, with a corresponding wavelength of \SI{1.5406}{\angstrom}.

\section{Results and discussion}

\subsection{XRD characterization}\label{subsec:XRD}

\begin{figure}[ht!]
\centering
\includegraphics[scale=0.21]{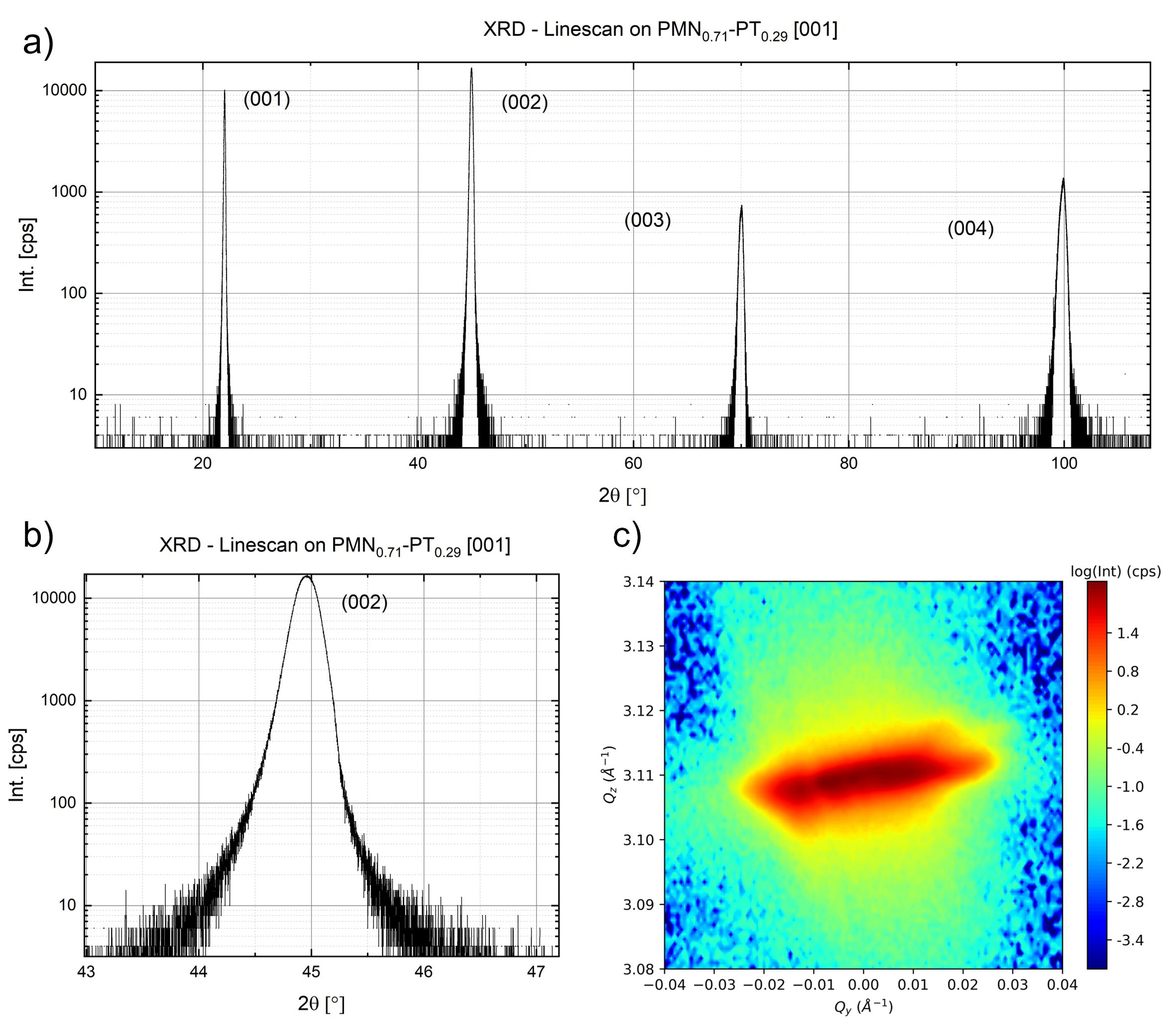}
\caption{Panel (a): a high resolution omega-2theta line-scan of an $\mathrm{PMN_{0.71}-PT_{0.29}}$ [001] substrate, several (00I)-PMN-PT Bragg peaks can be identified in the scan. Panel (b): zoomed-in regions around the (001) and (002) Bragg reflections scan. A broadening of the individual Bragg peaks ($\approx$ \SI{0.5}{\degree}) indicates a domain-like structure of the measured substrate with a certain distribution of lattice spacings (with a relative variation of the lattice constant  of $\approx 1\%$). Panel (c): reciprocal space map around the (002) reflection. The broadening in $Q_y$ can be attributed to a tilt distribution of the individual domains with respect to the [001] crystalline direction (at $Q_y=0$) and the broadening along $Q_z$ can again attributed to slightly different lattice parameters present in the material.}
\label{fig:XRD_data}
\end{figure}

Figure \ref{fig:XRD_data} (panel (a)) shows a high-resolution omega-2theta scan performed on a single crystalline, [001]-oriented, PMN-PT substrate. The scan shows, as expected, a series of [00I] Bragg reflections due to the fact that this set of lattice planes is oriented parallel to the sample surface. In the zoomed-in region around the individual Bragg reflections, (panel (b) of the same figure), one can see a relatively large full width at half maximum (FWHM) of the individual Bragg peaks in the range of about \SI{0.5}{\degree} that is attributed to a more domain-like structure of the PMN-PT substrate than expected for a perfect single crystal. A reciprocal space map (RSM) around the (002) reflection is shown in figure (panel (c)) of Figure \ref{fig:XRD_data}, where a clear broadening of the peak along the Qy direction indicates the presence of domains with a tilt distribution ($\approx$ \SI{0.7}{\degree}), whereas the broadening along the Qz direction accounts for domains with slightly different lattice parameters. These results demonstrate that the PMN-PT substrates are not perfectly single crystalline but rather a distribution of domains with different tilt angles with respect to the [001] direction. This has an impact on the mechanical stability of the substrates, since presence of possible defects at domain boundaries can make the crystal more fragile and difficult to process. More details about the substrates used in this work can be found in \cite{Ziss_2017_ComparisonofDifferentBondingTechniquesforEfficientStrainTransferUsingPiezoelectricActuators}.

\subsection{Ablation threshold, ablation rate and heating effects}\label{subsec:ablation}

We measured the ablation threshold for the PMN-PT crystals using the method of Liu \cite{Liu_1982_SimpleTechniqueforMeasurementsofPulsedGaussianBeamSpotSizes}.

\begin{figure}[ht!]
\centering
\includegraphics[scale=0.18]{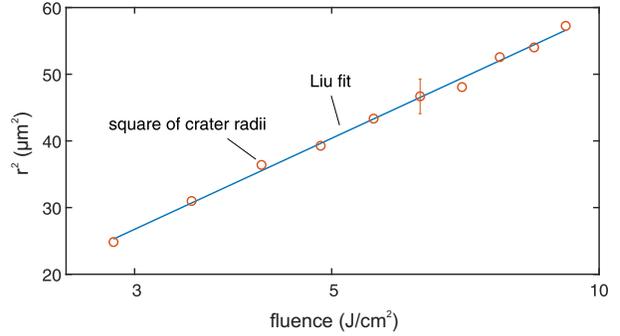}
\caption{Liu plot for the TRS crystals; a representative error bar is shown on one of the experimental data points. The value of the ablation threshold resulting from the fit is \SI{1.10(7)}{\joule \per \square \centi \metre}.}
\label{fig:Liu_plots}
\end{figure}

In this method one plots the square of the radii of the ablation craters generated by laser pulses versus the logarithm of the pulse fluence; from these plots one obtains both beam radius and ablation thresholds with a linear fit. Liu plots for single-pulse ablation of PMN-PT are shown in Figure~\ref{fig:Liu_plots}; the measured value of the ablation threshold for the TRS crystals is \SI{1.10(7)}{\joule \per \square \centi \metre}.

This ablation threshold for PMN-PT is higher than the one found by Di Maio and co-authors for PZT ceramic for \SI{120}{\femto \second} pulses at the wavelength of \SI{800}{\nano \meter}  (\SI{0.38}{\joule \per \square \centi \metre})\cite{DiMaio_2012_UltrafastLaserAblationCharacteristicsofPZTCeramicAnalysisMethodsandComparisonwithMetals}.

\begin{figure}[ht!]
\centering
\includegraphics[scale=0.5]{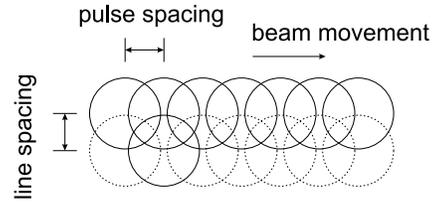}
\caption{Illustration of the laser cutting procedure. Each circle represents a laser pulse incident on the sample surface; the laser beam moves in sequence along parallel lines; we refer to the distance at which subsequent pulses hit the surface of the material as ``pulse spacing'' and to the separation between parallel lines as ``line spacing''.}
\label{fig:cutting_strategy}
\end{figure}

For both ablation and cutting we move the laser beam in sequence along parallel lines (see Figure \ref{fig:cutting_strategy}); in the following we will refer to the distance at which subsequent pulses hit the surface of the material as ``pulse spacing'' and to the separation between parallel lines as ``line spacing''.

The ablation rate per unit energy varies as a function of pulse energy, pulse spacing and repetition rate. The energy dependence is due to physical processes which are especially complicated for the case of dielectrics (see \cite{Rethfeld_2017_ModellingUltrafastLaserAblation} and \cite{Balling_2013_FemtosecondLaserAblationDynamicsofDielectricsBasicsandApplicationsforThinFilms});
for metals, where a considerable part of the absorption of the light pulses is realized through linear processes, it has been modelled by several authors  \cite{Nolte_1997_AblationofMetalsbyUltrashortLaserPulses,Neuenschwander_2014_SurfaceStructuringwithUltraShortLaserPulsesBasicsLimitationsandNeedsforHighThroughput} by simply considering that an optical pulse attenuates as it propagates inside the material and causes ablation if the fluence is larger than the ablation threshold; this model (which we call in the following ``logarithmic model'') predicts that there is an energy at which the ablation is most efficient and has been used for the modelling of ablation of glass as well \cite{Ben-Yakar_2004_FemtosecondLaserAblationPropertiesofBorosilicateGlass}. Despite its simplicity, the logarithmic model fits reasonably well our experimental ablation rate for acceptable values of the parameters, as we will discuss in the next paragraphs.

The pulse spacing and repetition rate dependencies, in general terms, result from a combination of heating and incubation effects. Placing pulses closer to each other both in time and space causes the sample to heat uniformly as each of the pulses releases some heat into the material and the dissipation is not fast enough to cool the material down before the arrival of the subsequent pulse (see for example the review of Finger et al. \cite{Finger_2018_HeatInputandAccumulationforUltrashortPulseProcessingwithHighAveragePower}); the heating of the material lowers the energy that must be deposited into it to reach evaporation and, as a consequence, lowers the ablation threshold and raises the ablation rate (see \cite{Ancona_2009_FemtosecondandPicosecondLaserDrillingofMetalsatHighRepetitionRatesandAveragePowers} for an example of ablation rate increase in metals).
An estimate of the cooling time precise enough to be useful is made difficult by the lack of an estimate of the energy penetration depth at fluences below the ablation threshold; the knowledge that the cooling time is of the order of microseconds (e.g. \cite{Sakakura_2007_HeatingandRapidCoolingofBulkGlassAfterPhotoexcitationbyaFocusedFemtosecondLaserPulse}) is insufficient to predict whether the onset of heat effects will be reached at, for example, \num{10} or \SI{50}{\kHz}.

Incubation is the decrease in the ablation threshold of a material with the increase of the number of laser pulses that are applied to a given spot \cite{Ashkenasi_1999_SurfaceDamageThresholdandStructuringofDielectricsUsingFemtosecondLaserPulsestheRoleofIncubation}; it can result in either an increase \cite{Byskov-Nielsen_2010_ShortPulseLaserAblationofMetalsFundamentalsandApplicationsforMicroMechanicalInterlocking} or a decrease \cite{Lenzner_1999_IncubationofLaserAblationinFusedSilicawith5FsPulses} in the ablation rate depending on the mechanism and on the values of the related changes of the physical properties of the material. We did not perform any experiment to quantify the effect of incubation in our case; however, it is sensible to expect them to be similar for all parameter sets: in our experiments different parameter sets are not distinguished by the number of pulses per unit area, what varies is the sequence of the pulse placement.

We studied ablation as a function of the pulse energy/pulse spacing/repetition parameters in the following way. For each material, we ablated two grids of \SI{0.5}{\milli \metre}-diameter circular pits using the ablation strategy illustrated in Figure \ref{fig:cutting_strategy}. In the first grid, we held the repetition rate fixed at  \SI{25}{\kilo \hertz} and varied energy (from about \num{0.4} to \SI{5.8}{\micro \joule}) and pulse spacing (from to \num{2} to \SI{10}{\micro \meter}), while in the second one we held the pulse spacing fixed (at  \SI{5}{\micro \meter}) and varied energy (again from about \num{0.4} to \SI{5.8}{\micro \joule}) and repetition rate (from to \num{12.5} to \SI{100}{\kilo \hertz}); in both cases we set the line spacing equal to the pulse spacing, so to have the surface uniformly covered with pulses after each pass of the laser. For each pit we chose the number of passes to reach a depth in the range of \num{10} to \SI{40}{\micro \metre} (obtaining a number of passes between 3 and 60); for each pass we chose randomly the direction of the laser beam movement.

\begin{figure}[ht!]
\centering
\includegraphics[scale=0.18]{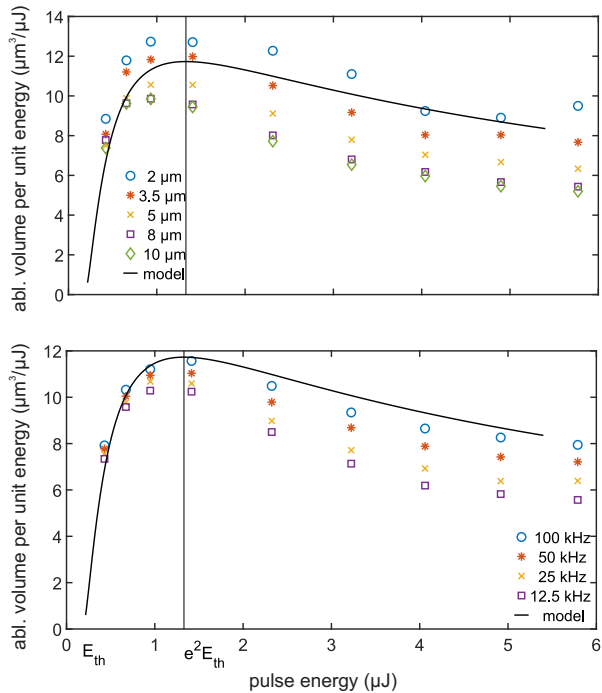}
\caption{Ablation rate for PMN-PT crystals.  Upper panel: pulse spacing dependence; lower panel: repetition rate dependence. Superimposed onto the experimental data is a qualitative fit to the logarithmic model for ablation; the parameters for both curves are fluence ablation threshold $\phi_{th} =  \SI{0.32}{\joule\per\cm\square}$ and energy penetration depth $\delta =  \SI{0.14}{\micro\meter}$. The threshold energy found by the model fit is marked in the abscissa as $\mathrm{E_{th}}$ and the corresponding predicted optimal pulse energy as $\mathrm{e^2 E_{th}}$.}
\label{fig:ablation_rate_02}
\end{figure}

We measured the depth of the pits using a profilometer (Dektak 8 Advanced Development Profiler); Figure~\ref{fig:ablation_rate_02} shows the ablated volume (calculated from the measured pit depths) per unit energy as a function of pulse energy. Ablation rate measurements performed on different days differed by about 10 \% in the high fluence range, while were reproducible to a few percent in the low fluence range (below approximately \SI{3}{\joule\per\square\cm}); the variability might be due exclusively to difference in laser parameters (e.g small differences in focusing) as different PMN-PT substrates measured on the same day gave results in good accord with each other. Superimposed on the data there is a qualitative fit to the logarithmic model, according to the following equation that takes into account the Gaussian profile of the laser beam \cite{Neuenschwander_2014_SurfaceStructuringwithUltraShortLaserPulsesBasicsLimitationsandNeedsforHighThroughput}:
\begin{equation}\label{eq:logarithmic_model}
\Delta V_E = \frac{1}{4}\pi {w_0}^2 \delta \log^2 \left( \frac{\phi_0}{\phi_{th}} \right)\frac{1}{E},
\end{equation}
where $\Delta V_E$ is the volume ablated per unit energy, $w_0$ the Gaussian beam radius, $\delta$ the energy penetration depth and $\phi_0$ and $\phi_{th}$ the fluence at the center of the beam and the ablation threshold fluence respectively, and $E$ is the pulse energy; one notices the sharp increase of the ablation efficiency next to the ablation threshold, the slower decrease for high pulse energy and the highest efficiency reached for an energy of approximately $e^2$ times the threshold for ablation.
The parameters for both qualitative fits are $\phi_{th} =  \SI{0.32}{\joule\per\cm\square}$ and $\delta =  \SI{0.14}{\micro\meter}$; the fluence threshold is lower than the one measured with the method of Liu for single pulse ablation and the energy penetration depth results from nonlinear absorption as PMN-PT is transparent in the visible spectrum \cite{Wan_2004_InvestigationonOpticalTransmissionSpectraof1xPbMg13Nb23O3xPbTiO3SingleCrystals};
the lower ablation threshold can be explained by the phenomenon of incubation (discussed in the preceding paragraphs) and the lower energy penetration depth with nonlinear absorption.
Despite the nonlinearity in the physical processes of the ablation, the simple logarithmic model fits reasonably well the shape of the curve of ablation rate versus energy using just two parameters.
The character of the ablation rate curve, and the presence of an optimum energy, are thus due to the logarithmic dependence of the ablation depth on the pulse fluence and on the Gaussian beam profile; the ablation rate grows as the square of the logarithm of the pulse fluence because both the ablation depth at points at which the fluence is above threshold and the area of the points which are above threshold grow as the logarithm of the pulse fluence.

\begin{figure}[ht!]
\centering
\includegraphics[scale=0.45]{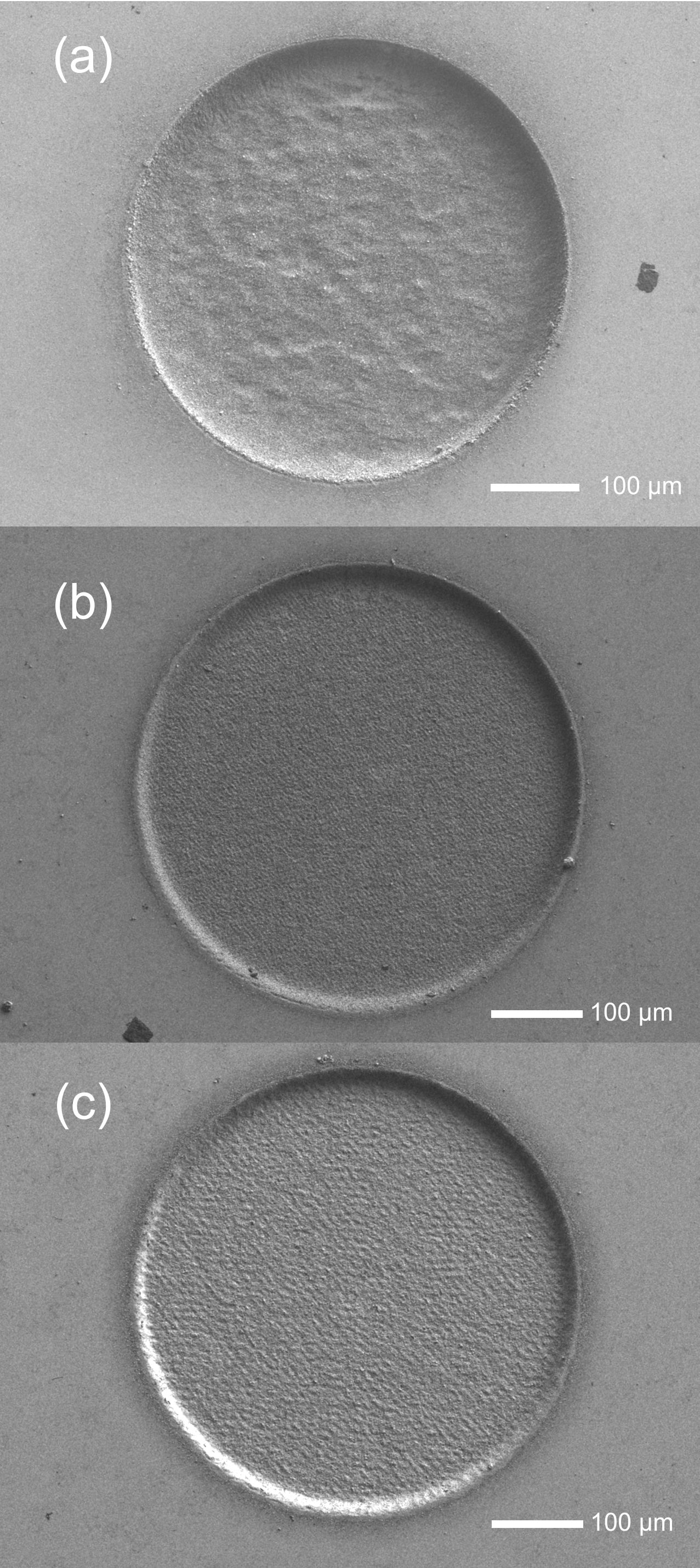}
\caption{Scanning electron microscope micrograph of ablated surfaces for various ablation parameters. Panel a): pulse spacing \SI{2}{\micro\meter}, energy \SI{5.5}{\micro\joule}; b): pulse spacing \SI{5}{\micro\meter}, energy \SI{5.5}{\micro\joule}; c): pulse spacing \SI{10}{\micro\meter}, energy \SI{5.5}{\micro\joule}. Repetition rate is \SI{25}{kHz} for all images.}
\label{fig:ablated_surfaces_electron_microscope_overview}
\end{figure}

Scanning electron micrographs of ablated surfaces are shown in Figure~\ref{fig:ablated_surfaces_electron_microscope_overview}; ablation with \SI{2}{\micro\meter} pulse spacing results in a rough surface topography (caused by irregular accumulation of debris) while the topography of the surface prepared with \SI{10}{\micro\meter} pulse spacing shows individual ablation craters and is for this reason rougher than the topography of the surface prepared with \SI{5}{\micro\meter} pulse spacing.

\begin{figure}[ht!]
\centering
\includegraphics[scale=0.5]{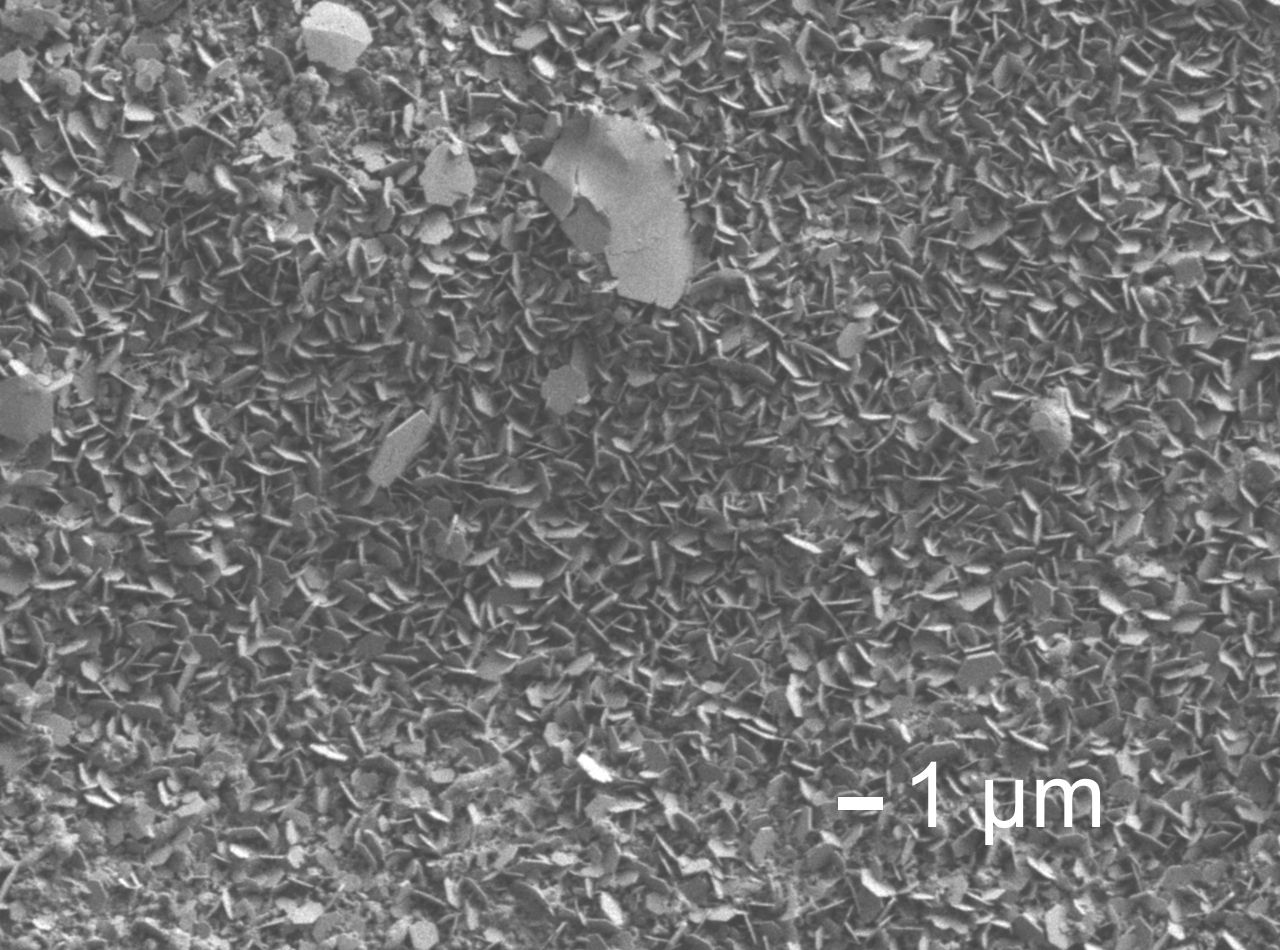}
\caption{Electron microscope micrograph of pit bottom showing hexagonal crystals; pulse spacing \SI{2}{\micro\meter}, energy \SI{5.5}{\micro\joule}, repetition rate \SI{25}{kHz}.}
\label{fig:ablated_surfaces_electron_microscope_bottom_crystals}
\end{figure}

Flat hexagonal crystals (nanoplatelets) appear at the bottom of many of the ablated pits for all processing parameters; we show them in Figure~\ref{fig:ablated_surfaces_electron_microscope_bottom_crystals} for the case of \SI{2}{\micro\meter} pulse spacing, \SI{5.5}{\micro\joule} energy and \SI{25}{\kHz} laser repetition rate. The origin of this interesting phenomenon is at present unknown and may deserve further investigation.

\begin{figure}[ht!]
\centering
\includegraphics[scale=0.50]{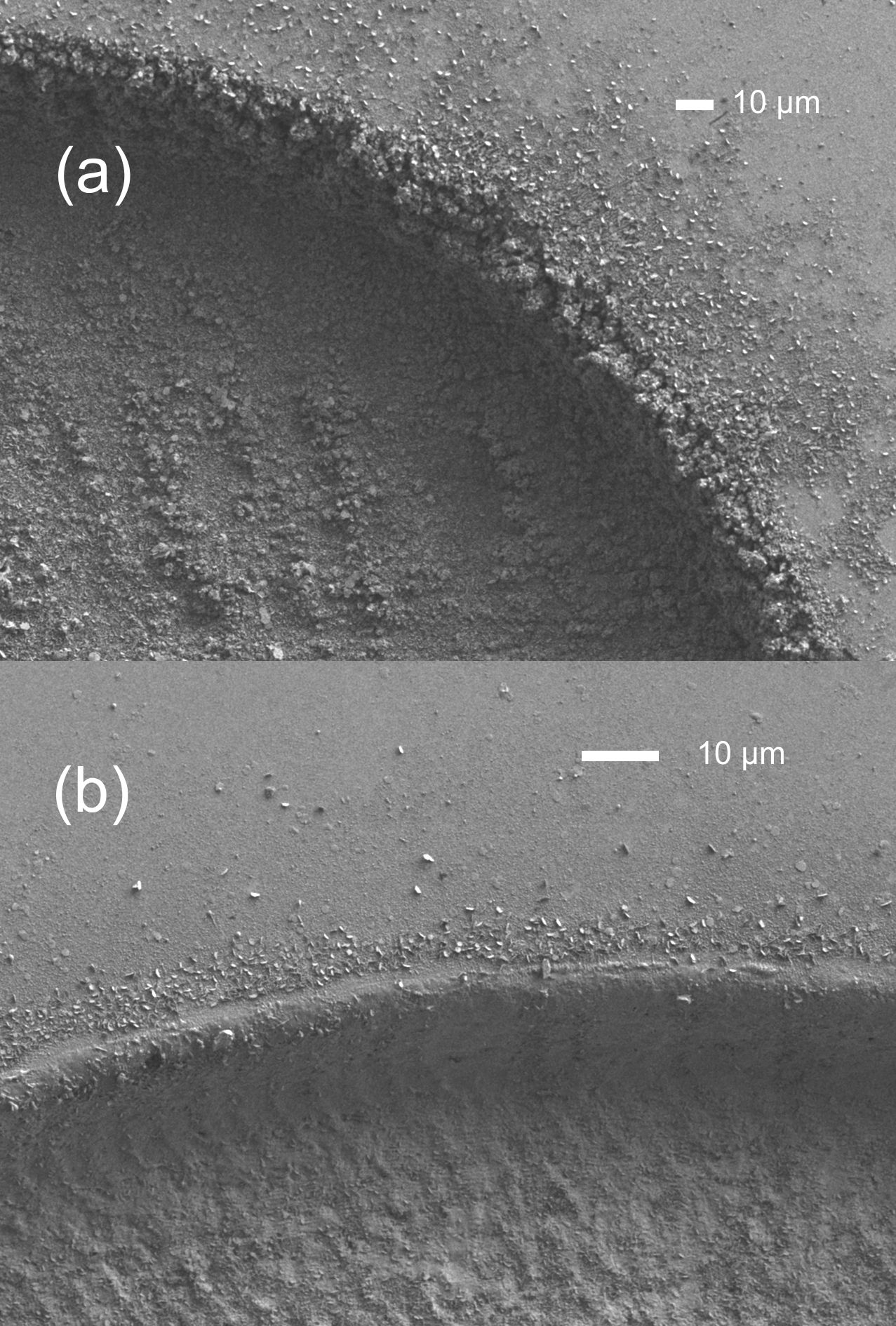}
\caption{Scanning electron microscope micrograph of pit edges. Panel a): pulse spacing \SI{2}{\micro\meter}, energy \SI{6.6}{\micro\joule}; b): pulse spacing \SI{5}{\micro\meter}, energy \SI{6.6}{\micro\joule}. Rpetition rate it \SI{25}{kHz} for both images.}
\label{fig:ablated_surfaces_electron_microscope_edge}
\end{figure}

The edges of the ablated pits (shown in Figure~\ref{fig:ablated_surfaces_electron_microscope_edge}) are generally good except for the case of pulse spacing equal to \SI{2}{\micro\meter}, where they are irregular; the jagged borders appear to result from redeposition of ablated material, although the debris might hide minor chipping (that is, separation of small fragments of materials) of the edges.
A good edge quality is more difficult to achieve for cutting than for the ablation of an extended region, as cutting is performed by moving the laser in sequence along a few parallel lines; in the case of surface ablation we arranged the path of the laser so that there is only one pulse per laser line next to the edge, so the edges have more time to cool down between subsequent pulses, while for the cuts all of the pulses are next to the edge. We examine this case more in detail in Subsection~\ref{subsec:edge_wall_quality}.

In conclusion, with green pulses of the duration of \SI{380}{\fs} and a beam radius of \SI{6}{\micro\meter} it is possible to ablate approximately \SI{2.5}{\mm \cubed \per \minute \per \watt} of PMN-PT without heat effects and with limited redeposition of ablated material on the surface; heating starts to affect ablation strongly with pulse spacing below \SI{5}{\micro\meter} and with pulse repetition rate above \SI{25}{\kilo\hertz}.

\subsection{Edge and wall quality}\label{subsec:edge_wall_quality}

The cutting of crystals is realized by placing several rows of pulses next to each other (Figure \ref{fig:cutting_strategy}); repeating this procedure with a Gaussian beam results in a V-shaped groove that, when deep enough, separates two parts of the crystal.
Two issues arise. The first issue is that cracks appear more easily than in the case of surface ablation; the second issue is that for a Gaussian beam the groove walls are inclined and become steeper as the pulse energy grows.

Cracking in the ablation with ultrashort (ps) laser pulses has been studied by Sun and co-authors for the case of glass \cite{Sun_2016_DamageMorphologyandMechanisminAblationCuttingofThinGlassSheetswithPicosecondPulsedLasers}. According to their study, cracks are caused by the thermal shock of cooling;  high temperatures in the case of ultrashort pulses are caused by the accumulation of energy released in the material by each of the pulses in a pulse train.

It is reasonable that in our experiments cracks should appear in cutting rather than surface ablation since cutting and ablation are realized with different scanning strategies; in the case of cutting one is placing several lines of pulses adjacent to each other, so that all of the pulses are close to the edge, while in the case of ablation, as illustrated in the previous section, laser scanning proceeds in random directions with respect to the edges so that usually just a few consecutive pulses are placed next to an edge. In our experiments, as we will discuss in the following paragraphs, we have seen that cracking appears at the relatively low repetition rate of \SI{25}{\kilo\hertz} if the pulse energy is too high.
As an additional note, in the mechanism of cracking there may be other components in addition to cooling shocks: cracks have been observed in glass after three ps pulses delivered at the frequency of \SI{1}{\hertz} \cite{Herman_1999_LaserMicromachiningofTransparentFusedSilicawith1PsPulsesandPulseTrains} and under these conditions heat accumulation can be excluded.

\begin{figure}[ht!]
\centering
\includegraphics[scale=0.40]{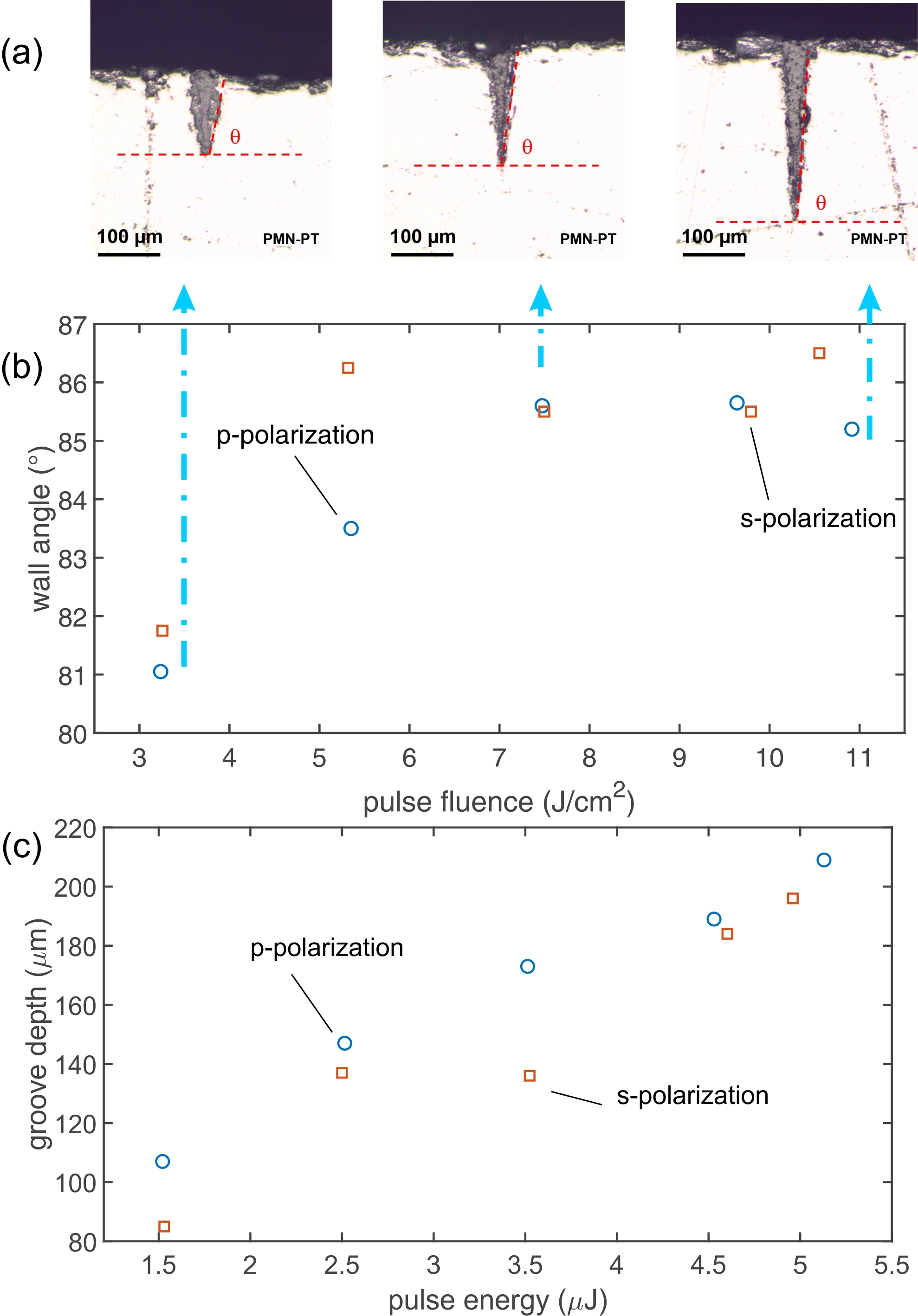}
\caption{Panels (a) and (b): wall angle as a function of pulse fluence for p- and s-polarized light. Panel (c): groove depth as a function of pulse energy for p- and s-polarized light. The grooves were cut till saturation; beam radius is \SI{5.5}{\micro\meter}, pulse spacing is \SI{5}{\micro\meter} and laser repetition rate is \SI{25}{\kilo\hertz}. The cuts are realized by placing 5 parallel lines of pulse at the distance of \SI{5}{\micro\meter} from each other.}
\label{fig:groove_depth}
\end{figure}

A qualitative explanation for the V-shape of the groove (as for example discussed in \cite{Fornaroli_2013_DicingofThinSiWaferswithaPicosecondLaserAblationProcess}) is that after the first laser pass the ablation is not uniform, but is deeper at the center of the beam, where the fluence is higher; subsequent passes of the laser ablate on an inclined surface, so that the pulse fluence projected onto the surface becomes lower. Ablation stops when the projected fluence is equal to the threshold fluence, resulting in V-shaped grooves (see panel (a) of Figure~\ref{fig:groove_depth} for examples of grooves cut in PMN-PT crystals); an attempt to explain quantitatively the steepening of the walls and the consequent stopping of ablation has been made by V\'azquez de Aldana and coworkers for the case of fused silica including the details of propagation of the light pulses in the developing groove \cite{VazquezdeAldana_2006_SaturationofAblationChannelsMicroMachinedinFusedSilicawithManyFemtosecondLaserPulses}. In the following we refer to the stopping of ablation when the groove walls have reached a limit angle as ``saturation".
A steep wall allows narrower cuts (fewer rows of pulses need to be placed alongside each other in order to cut through the crystal thickness), so that cutting is quicker and smaller features can be cut.
Improving edge quality by lowering the fluence then is achieved at expense of the wall steepness.

In order to investigate these issues we carried out a cutting experiment. On a PMN-PT crystal we placed five parallel lines of pulses, each approximately \SI{3.5}{\milli\meter}  long, at the distance of \SI{5}{\micro\meter} from each other (as in the usual cutting procedure) and we varied the pulse energy for two different polarizations: perpendicular to the cutting lines (and so p-polarized with respect to the walls of the resulting groove) and parallel to the cutting line (and so s-polarized); the \SI{5.5}{\micro\meter} beam radius was placed at the crystal surface.

Since from our ablation experiments (Section \ref{subsec:ablation}) we determined that heating effects are limited for pulse spacings of \SI{5}{\micro\meter} and laser repetition rates of \SI{25}{\kilo\hertz} and smaller, we fixed \SI{5}{\micro\meter} pulse spacing and \SI{25}{\kilo\hertz} repetition rate as processing parameters for all cutting experiments. We executed 500 passes to be sure that in all cases we cut till saturation; with five cutting lines none of the grooves cut through the \SI{300}{\micro\meter}-thick crystal so that for each parameter set it is possible to measure the maximum cutting depth.

We measured the resulting cut depth and we examined the cut edge quality both at the optical and the scanning electron microscope. For the measurement of the cut depth, we embedded the processed crystal in resin and we sectioned them so that the measurements could be carried out with an optical microscope. For this data set and the following data set on surface cracks the uncertainty on the knowledge of the beam radius is larger than in the case of the ablation rate experiments; this could possibly lead to an underestimation of the cutting depths and of an overestimation of the maximum energy at which no cracks appear (as any imprecision in the placement of the focus can only lead to larger beam radii).

A graph of the groove depths is shown in panel (c) of Figure~\ref{fig:groove_depth}. For a given energy, p-polarized light generates deeper cuts than s-polarized light (a known fact in general, see for example \cite{Venkatakrishnan_2002_TheEffectofPolarizationonUltrashortPulsedLaserAblationofThinMetalFilms}). This can be explained with a larger absorption of p-polarized with respect to s-polarized light at the inclined walls; the price to pay is that with  p-polarized light cracks and chipping form more readily than with  s-polarized light.
We do not possess a statistics on this very time-consuming experiment, but we can assume that the scatter of the results would be at least as high as the one observed for the ablation rates.

The angles that the groove walls form with the horizontal plane (in the following ``wall angles'') are shown in panel (b) of Figure~\ref{fig:groove_depth} (with the accompanying panel (a) showing representative grooves) as a function of pulse fluence (the most relevant parameter for the wall angle).

\begin{figure}[ht!]
\centering
\includegraphics[scale=0.32]{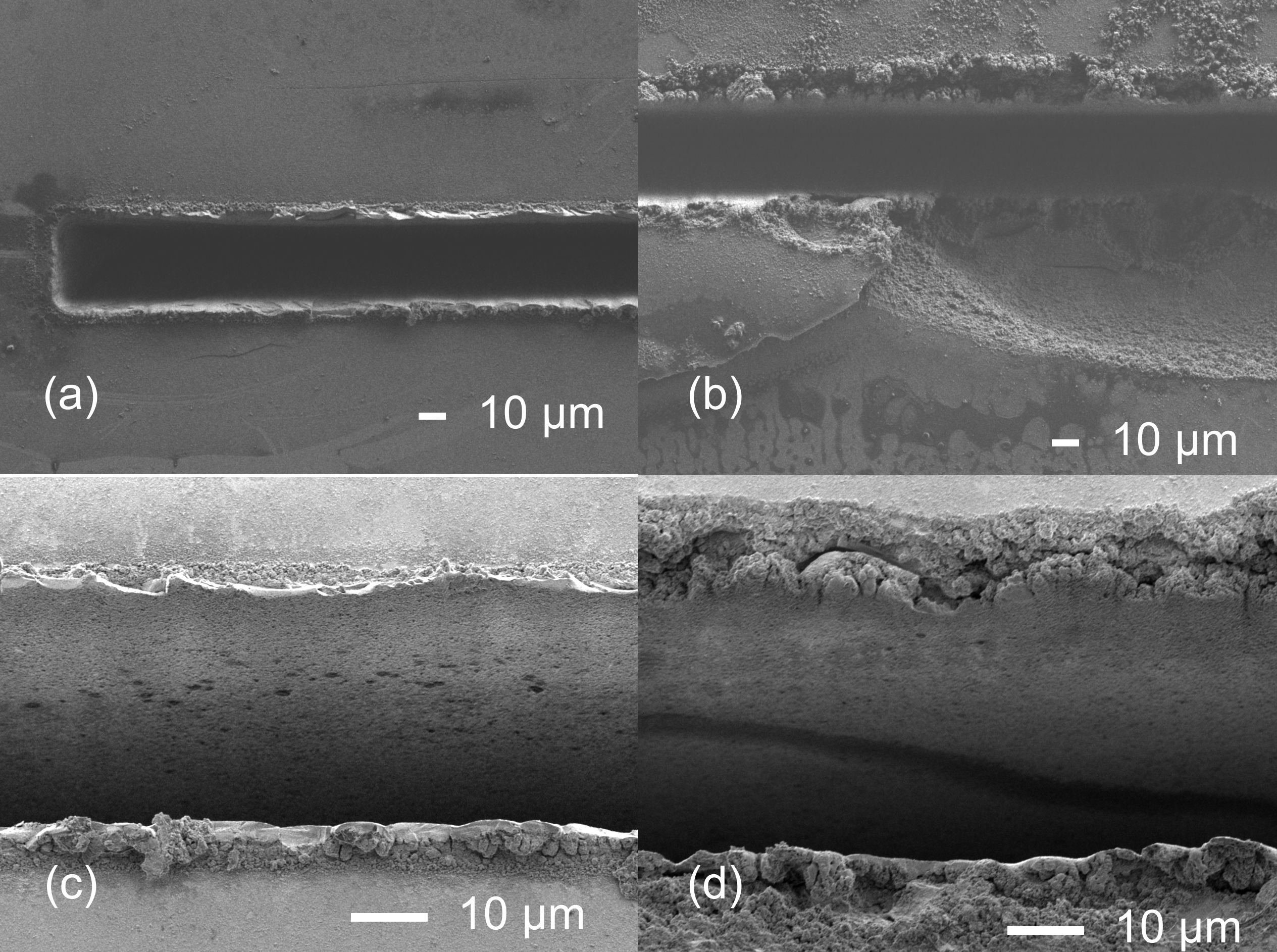}
\caption{Scanning electron micrographs of cut edge (top row) and cut wall (bottom row) for cuts performed with p-polarized light. Beam radius at waist is \SI{5.5}{\micro\meter}, pulse spacing is \SI{5}{\micro\meter} and laser repetition rate is \SI{25}{\kilo\hertz}; the cuts are realized by placing 5 parallel lines of pulses at the distance of \SI{5}{\micro\meter} from each other; pulse energy is \SI{3.5}{\micro\joule} (left column) and \SI{5.7}{\micro\joule} (right column); the beam waist is set at the surface. The cut edge at \SI{5.7}{\micro\joule} is cracked and chipped, while at \SI{3.5}{\micro\joule} the cracking is minimal; the cut walls are smooth in both cases.}
\label{fig:edge_quality_electron_microscope}
\end{figure}

\begin{figure}[ht!]
\centering
\includegraphics[scale=0.30]{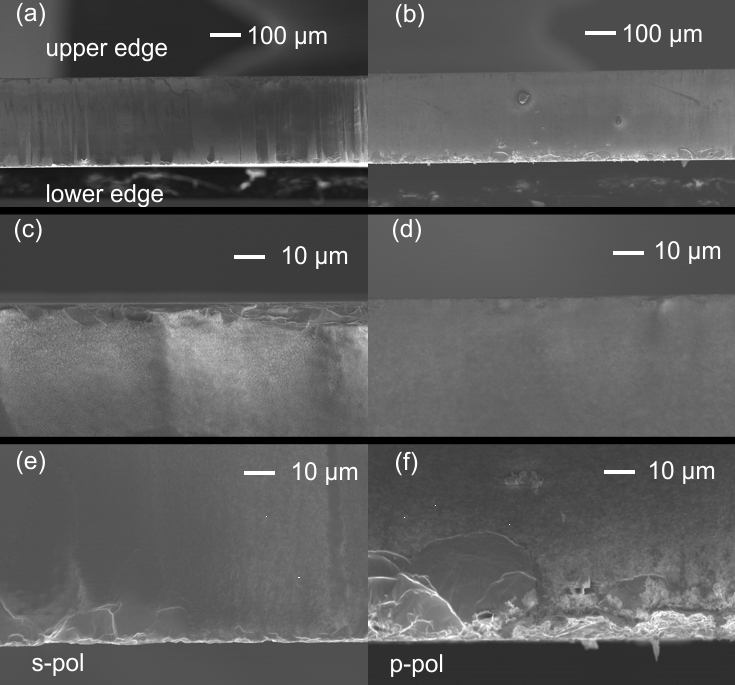}
\caption{Side walls of a complete cut of a \SI{300}{\micro\meter}-thick PMN-PT substrate realized with pulses of \SI{5.5}{\micro\joule}, repetition rate of \SI{25}{\kHz} and the pulse and line spacing of \SI{5}{\micro\meter} for 30 parallel lines and 500 laser passes; the laser focus is placed in the middle of the substrate. Left column, s-polarization, right column, p-polarization; panels (a) and (b): overview, panels (c) and (d): detail of upper edge, panels (e) and (f): detail of lower edge. The walls are smooth with a small amount of chipping at the wall bottom.}
\label{fig:full_walls}
\end{figure}

\begin{figure}[ht!]
\centering
\includegraphics[scale=0.45]{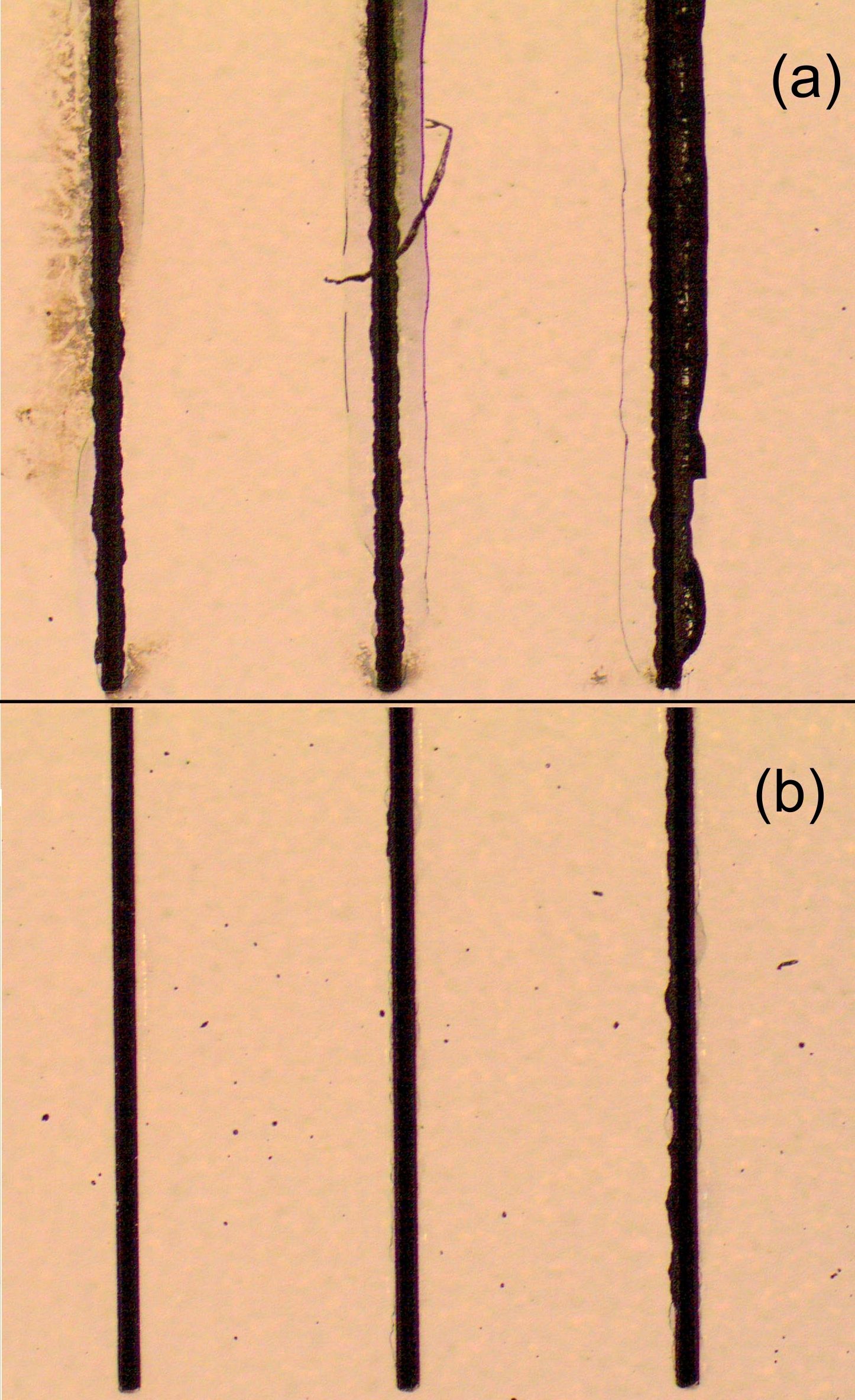}
\caption{Optical microscope images of cuts performed in PMN-PT crystals under different conditions. Panel (a): beam waist at the surface, p-polarized light; panel (b): beam waist at the surface, s-polarized light. For all images, the beam radius at waist is \SI{6}{\micro\meter}, the pulse spacing is \SI{5}{\micro\meter} and the laser repetition rate is \SI{25}{\kilo\hertz}; the cuts are realized by placing 5 parallel lines of pulse at the distance of \SI{5}{\micro\meter} from each other; pulse energy is in each panel 4.5, 5.5 and \SI{6.3}{\micro\joule} for left, middle and right cut respectively. For s-polarized light and the energy of \SI{6.3}{\micro\joule} cracks are observed only in some of the cuts (not shown in the figure).}
\label{fig:edge_quality_optical_microscope}
\end{figure}

A high pulse fluence leads to steeper walls for both light polarizations (advantageous, as smaller details can be cut in the piezocrystals) but too high a fluence results in chipping and cracking of the cut edges (see Figures~\ref{fig:edge_quality_electron_microscope} and \ref{fig:edge_quality_optical_microscope}). The walls of the cut are smooth in this material next to the top (see Figure \ref{fig:edge_quality_electron_microscope}). We analysed the full wall by cutting completely a \SI{300}{\micro\meter}-thick substrate with pulses of \SI{5.5}{\micro\joule} energy using the repetition rate of \SI{25}{\kHz} and the pulse and line spacing of \SI{5}{\micro\joule} for 30 parallel lines and 500 laser passes; we placed the focus below the surface to avoid cracking (as we discuss in more detail in the following paragraphs). The resulting walls, shown in Figure \ref{fig:full_walls} are smooth with a small amount of chipping at the wall bottom.

The edge quality has a strong dependence on the polarization of the incident light. Cuts performed with s-polarized light chip and crack only at the highest pulse energy which we used in the experiments (about \SI{6.3}{\micro\joule}) while cutting with p-polarized light results in cracking already at \SI{3.5}{\micro\joule}; Figure \ref{fig:edge_quality_optical_microscope} shows a comparison of edge quality for p- and s-polarization cutting. Combining the data on the edge quality with the data on the cut depth it would seem that cutting with s-polarized light is more convenient than cutting with p-polarized light, as it is possible to cut deeper given an acceptable edge quality; with a fixed waveplate the polarization with respect to the cut walls cannot be chosen if one is cutting a closed contour and one could attempt to cut with circularly polarized light as a compromise. In the literature several methods to tailor the light polarization, which in our case would afford more control on the quality of cuts with different orientations, have been studied; \cite{Jin_2013_DynamicModulationofSpatiallyStructuredPolarizationFieldsforRealTimeControlofUltrafastLaserMaterialInteractions,Jofre_2012_FastBeamSteeringwithFullPolarizationControlUsingaGalvanometricOpticalScannerandPolarizationController} pursued the road of dynamically changing the polarization while \cite{Torres_2013_InfluenceofLaserBeamPolarizationonLaserMicroMachiningofMolybdenum} experimented with circular, radial  and azimuthal polarization, which are insensitive to the direction of the cut (radial and azimuthal polarizations approximate for a cutting process respectively p- and s-polarization states). A practical solution for a \SI{300}{\micro\meter}-thick crystal with our \SI{7}{\micro\meter} radius at the waist, that allows us to circumvent the issue of polarization, is placing the beam waist at approximately the bottom of the crystal (a precise placement is irrelevant and is made impossible by the fact that the substrates are coated with photoresist to protect the surface from the fabrication dust); in this case the quality of the edges is good up to relatively high pulse energy independently of polarization, allowing one to disregard polarization issues in practical fabrication. Using this placement of the focus we were able to use a pulse energy of \SI{5.5}{\micro\joule} and still obtain a good edge quality; as an added advantage, with this placement of the focus one avoids the readjustment of the beam focus with the progression of cutting. Note that when ablating with pulse energies sufficiently higher than the threshold small focal shifts influence just weakly the ablation rate and may even lead to its increase \cite{Chen_2018_UltrafastZScanningforHighEfficiencyLaserMicroMachining}. We summarize practical device fabrication further in Section \ref{sec:conclusions}.

In conclusion, a good compromise between processing speed and surface/edge quality is a pulse energy of \SI{5.5}{\micro\joule} with a beam waist of \SI{7}{\micro\meter}, a repetition frequency of \SI{25}{\kilo\hertz} and the placement of the beam waist below the surface of a thick substrate; with these parameters one is able to ablate approximately \SI{0.3}{\mm \cubed \per \minute} of PMN-PT. With these parameters it is necessary to place the beam waist under the surface of the PMN-PT substrate in order to obtain crack-free edges. As a compensation for the reduced fluence at the surface, one is able to cut substrates of \SI{300}{\micro\meter} (and over) of thickness without adjustment of the beam focus.

\begin{figure}[ht!]
\centering
\includegraphics[width = \linewidth]{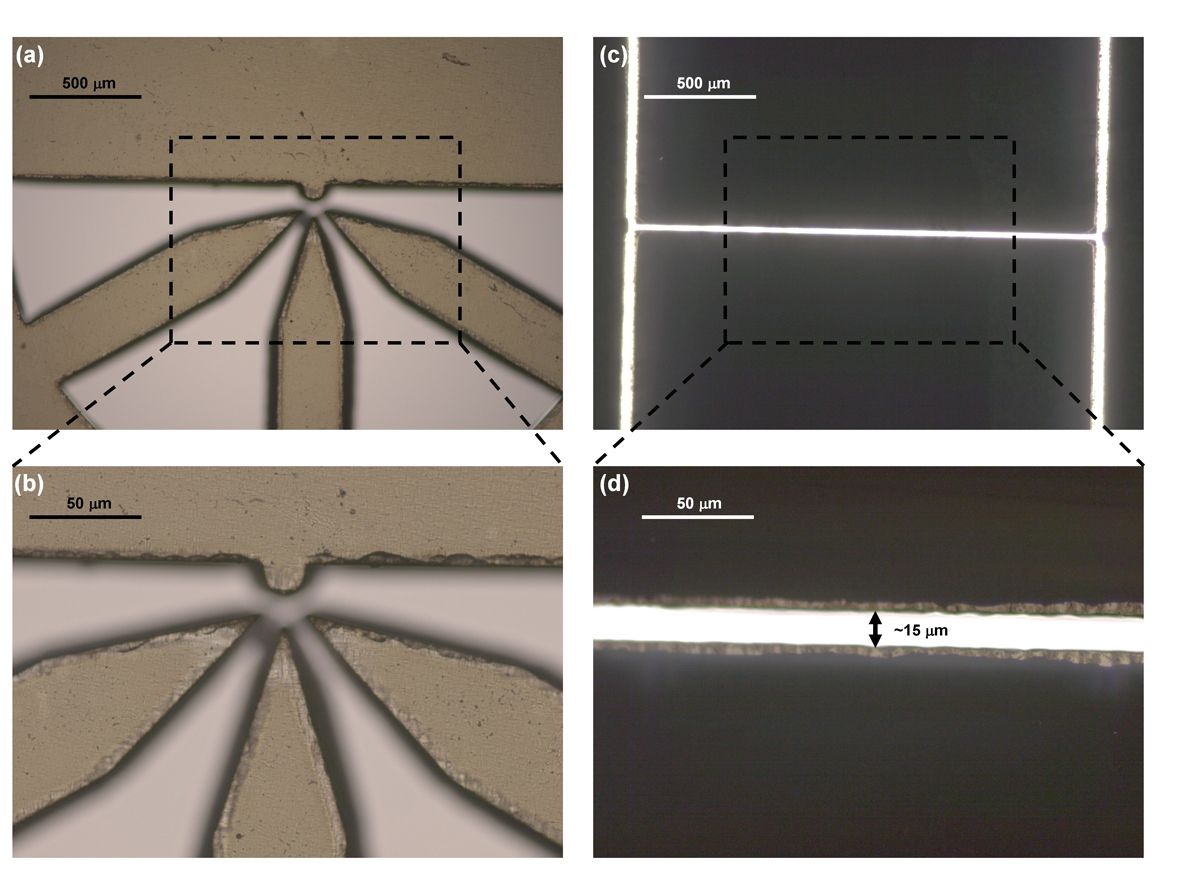}
\caption{Optical images of micro-machined 200 and \SI{300}{\micro\meter} thick PMN-PT substrates. Panel (a): a design featuring three legs with features in the micrometric scale fabricated in a \SI{300}{\micro\meter} thick substrate. Panel (b): a magnified picture of (a), where the gaps between the fabricated legs are as small as \SI{10}{\micro\meter}. Panels (c) and (d): optical microscope images obtained in transmitted illumination of a H-shape feature, fabricated in a \SI{300}{\micro\meter} substrate, with gaps of the order of \SI{15}{\micro\meter}. Note that the size of the gaps decreases from top to bottom due to the inclination of the cut walls.}
\label{fig:fabricated_devices}
\end{figure}

In figure \ref{fig:fabricated_devices} we show optical images of 300- and 200-micrometers-thick PMN-PT substrates cut with complex shapes. These substrates were cut with the optimized conditions described in the preceding paragraph in order to have sharp and well-defined edges. Electrical contacts have been fabricated on both sides of the piezoelectric substrate and a voltage ramp was applied in several cycles to further check the performance of the fabricated micro-machined substrates. We did not observe any evolution of cracks and we found that the electrical response of the devices was stable. We believe that the fabrication method presented in this work opens up new possibilities for the development of advanced piezoelectric devices where precise micro-scale feature definition is mandatory.

\section{Conclusions} \label{sec:conclusions}

Ultrafast laser microfabrication is a flexible and rapid way to obtain devices from PMN-PT crystals. A good compromise between processing speed and surface/edge quality is a pulse energy of \SI{5.5}{\micro\joule} with a beam waist of \SI{7}{\micro\meter}, a repetition frequency of \SI{25}{\kilo\hertz} and the placement of the beam waist below the surface of a thick substrate; with these parameters one is able to ablate approximately \SI{0.3}{\mm \cubed \per \minute} of PMN-PT and mm\nobreakdash-sized devices with deep (\SI{300}{\micro\meter}) structures and gaps smaller than \SI{50}{\micro\meter} can be obtained in under 30 minutes. The fabricated devices were tested after evaporation of gold electrodes by cycling 10 times voltage ramps from \SI{0}{\volt} to \SI{200}{\volt}. No visible degradation (evolution of cracks) was observed on the machined crystals after this process. Hence, such micro-machined crystals can be successfully employed for actuating applications as demonstrated in some of our recent works \cite{Martin-Sanchez_2018_StrainTuningoftheOpticalPropertiesofSemiconductorNanomaterialsbyIntegrationontoPiezoelectricActuators,Martin-Sanchez_2016_ReversibleControlofinPlaneElasticStressTensorinNanomembranes,Trotta_2016_WavelengthTunableSourcesofEntangledPhotonsInterfacedwithAtomicVapours,Huber_2018_StrainTunableGaAsQuantumDotaNearlyDephasingFreeSourceofEntangledPhotonPairsonDemand,Yuan_2018_UniaxialStressFlipstheNaturalQuantizationAxisofaQuantumDotforIntegratedQuantumPhotonics}.

\begin{acknowledgements}
The authors thank Johann Zehetner for precious indications, Thomas Auer for technical support with the sectioning of the crystals and Stephan Kasemann for electron microscope micrographs.
The work was supported financially by the European Union Seventh Framework Program 209 (FP7/2007-2013) under Grant Agreement No. 601126 210 (HANAS), the \begin{otherlanguage*}{ngerman}AWS Austria Wirtschaftsservice, PRIZE Programme,\end{otherlanguage*} under Grant No. P1308457,the European Research Council (ERC) under the European Unions Horizon 2020 research and innovation programme (SPQRel, Grant Agreement No. 679183), and the Christian Doppler Gesellschaft under the``Josef Ressel Zentrum f{\"u}r Materialbearbeitung mit ultrakurz gepulsten Laserquellen''. J.M.-S. acknowledges support from the Government of the Principality of Asturias through the Clarín Programme and a Marie Curie-COFUND grant (PA-18-ACB17-29).
\end{acknowledgements}

\end{document}